\documentclass{article}
\usepackage{spconf}
\usepackage{graphicx}
\begin{document}
\title{Attention-Constrained Inference for Robust Decoder-Only Text-to-Speech}
\name{Hankun Wang$^1$, Chenpeng Du$^1$, Yiwei Guo$^1$, Shuai Wang$^2$, Xie Chen$^1$, Kai Yu$^1$}

\address{
  $^1$MoE Key Lab of Artificial Intelligence, AI Institute;\\X-LANCE Lab, Department of Computer Science and Engineering,\\Shanghai Jiao Tong University, Shanghai, China \\
  $^2$Shenzhen Research Institute of Big Data, The Chinese University of Hong Kong, Shenzhen, China}

\maketitle              % typeset the header of the contribution
\begin{abstract}
Recent popular decoder-only text-to-speech models are known for their ability of generating natural-sounding speech. However, such models sometimes suffer from word skipping and repeating due to the lack of explicit monotonic alignment constraints. In this paper, we notice from the attention maps that some particular attention heads of the decoder-only model indicate the alignments between speech and text. We call the attention maps of those heads Alignment-Emerged Attention Maps (AEAMs). Based on this discovery, we propose a novel inference method without altering the training process, named Attention-Constrained Inference (ACI), to facilitate monotonic synthesis. It first identifies AEAMs using the Attention Sweeping algorithm and then applies constraining masks on AEAMs. Our experimental results on decoder-only TTS model VALL-E show that the WER of synthesized speech is reduced by up to 20.5\% relatively with ACI while the naturalness and speaker similarity are comparable.

\end{abstract}
\section{Introduction}
In the field of text-to-speech (TTS), numerous breakthroughs have been achieved based on deep neural networks \cite{Wang2017TacotronTE,ren2021fastspeech2,Wang2023NeuralCL}. TTS is a sequence-to-sequence task, where the text and speech possess a monotonic aligned nature, and the audio frame sequence is usually much longer than the text sequence.
Prior methods typically integrate an explicit duration module to fill this modality gap. These models obtain the target durations via conducting traditional alignment algorithms either externally \cite{ren2021fastspeech2,pmlr-v139-popov21a-gradtts,guo2023voiceflow} or internally \cite{kim2020glow,pmlr-v139-kim21f-vits}, and feed them as model input for training. 
At the training stage, using the target durations, the encoded text sequence is expanded to the length of speech frames and then serves as the decoder inputs. Meanwhile, the duration module is trained 
to predict the target durations. At the inference stage, the predicted durations are employed to expand the encoded text sequence for further decoding. In summary, monotonic alignment constraints are explicitly involved in this framework to avoid robustness issues like word skipping, repeating, or mispronouncing. 

\begin{figure*}
    \centering
    \begin{subfigure}[t]{0.16\textwidth}
        \centering
        \includegraphics[width=0.5\linewidth]{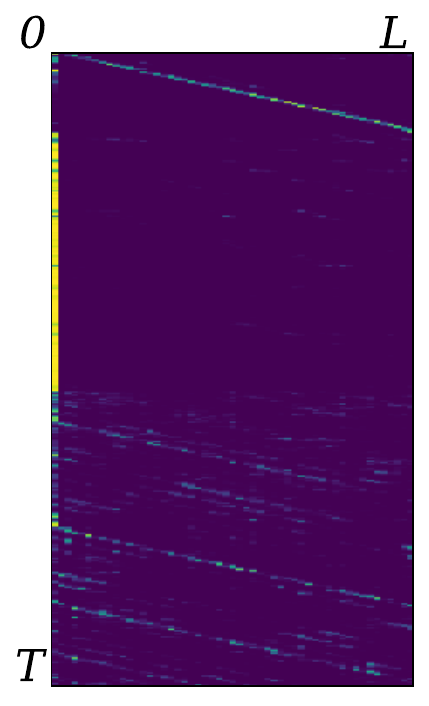}
        \caption{Head 1 of layer 1}
    \end{subfigure}
    \begin{subfigure}[t]{0.16\textwidth}
        \centering
        \includegraphics[width=0.5\linewidth]{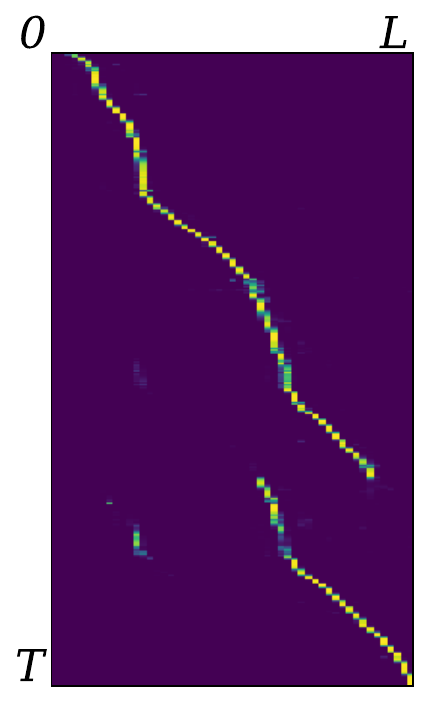}
        \caption{Head 4 of layer 2}
    \end{subfigure}
    \begin{subfigure}[t]{0.16\textwidth}
        \centering
        \includegraphics[width=0.5\linewidth]{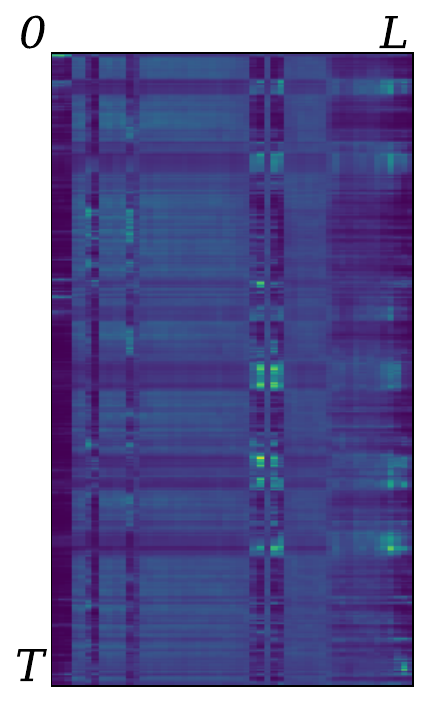}
        \caption{Head 3 of layer 3}
    \end{subfigure}
    \begin{subfigure}[t]{0.16\textwidth}
        \centering
        \includegraphics[width=0.5\linewidth]{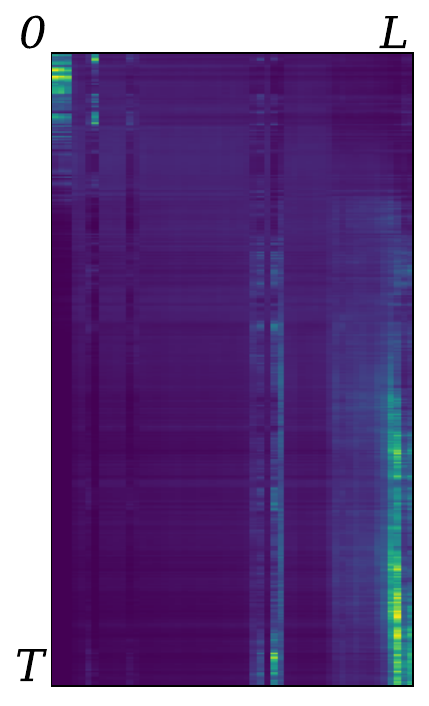}
        \caption{Head 3 of layer 4}
    \end{subfigure}
    \begin{subfigure}[t]{0.16\textwidth}
        \centering
        \includegraphics[width=0.5\linewidth]{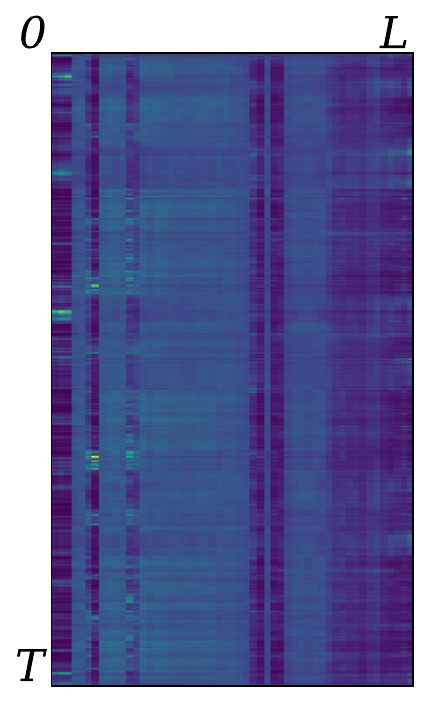}
        \caption{Head 1 of layer 5}
    \end{subfigure}
    \begin{subfigure}[t]{0.16\textwidth}
        \centering
        \includegraphics[width=0.5\linewidth]{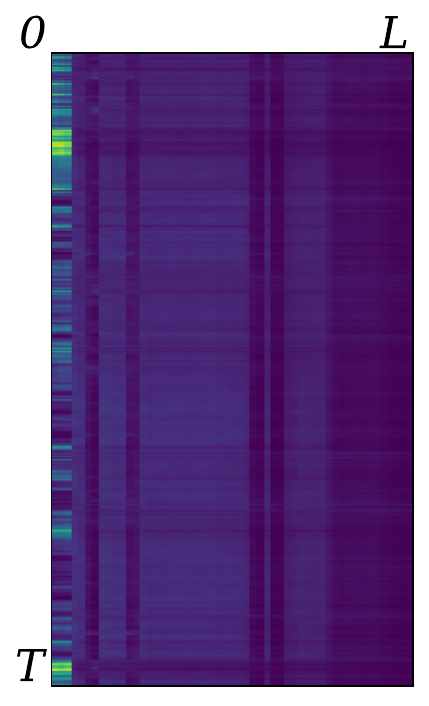}
        \caption{Head 4 of layer 6}
    \end{subfigure}
    \\ \vspace{0.5em}
    \begin{subfigure}[t]{0.16\textwidth}
        \centering
        \includegraphics[width=0.5\linewidth]{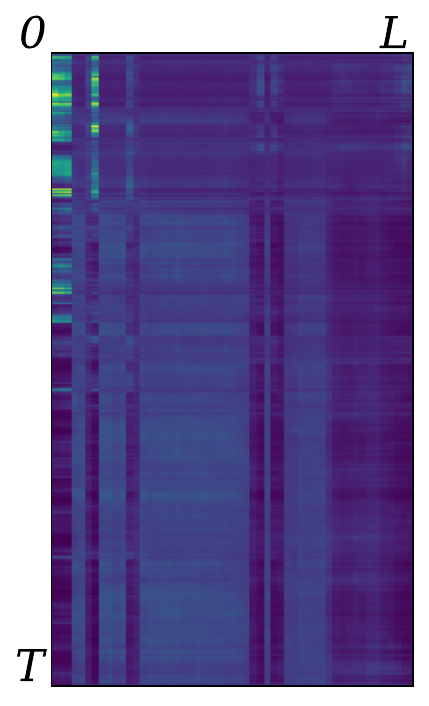}
        \caption{Head 8 of layer 7}
    \end{subfigure}
    \begin{subfigure}[t]{0.16\textwidth}
        \centering
        \includegraphics[width=0.5\linewidth]{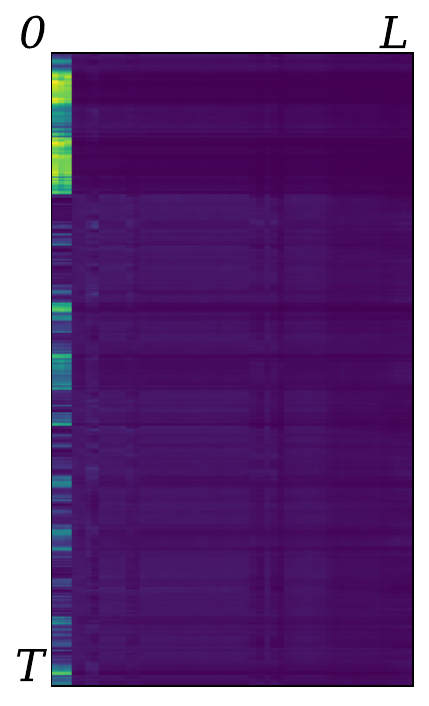}
        \caption{Head 8 of layer 8}
    \end{subfigure}
    \begin{subfigure}[t]{0.16\textwidth}
        \centering
        \includegraphics[width=0.5\linewidth]{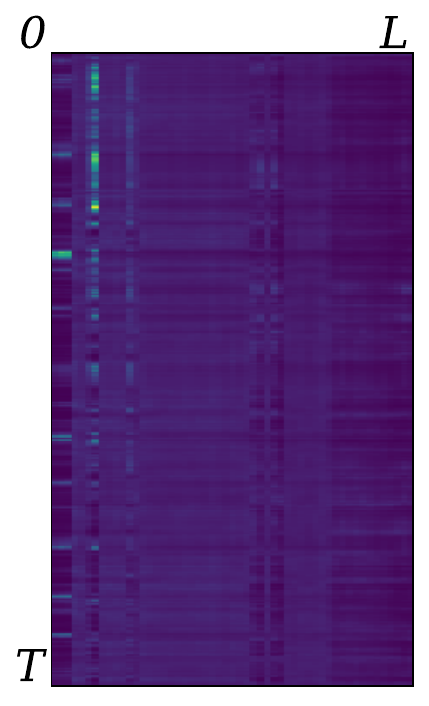}
        \caption{Head 13 of layer 9}
    \end{subfigure}
    \begin{subfigure}[t]{0.16\textwidth}
        \centering
        \includegraphics[width=0.5\linewidth]{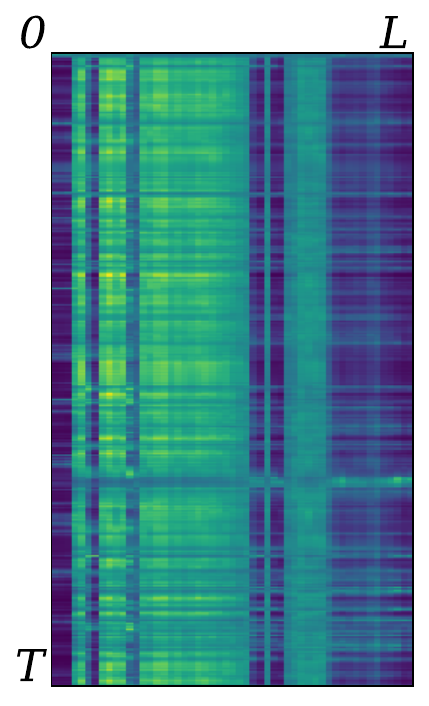}
        \caption{Head 12 of layer 9}
    \end{subfigure}
    \begin{subfigure}[t]{0.16\textwidth}
        \centering
        \includegraphics[width=0.5\linewidth]{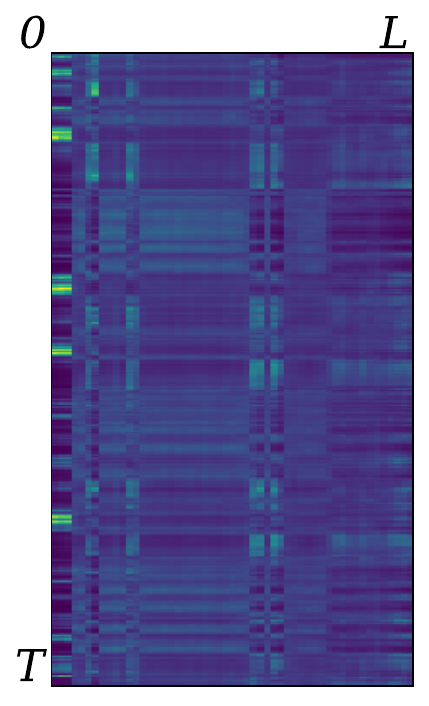}
        \caption{Head 8 of layer 11}
    \end{subfigure}
    \begin{subfigure}[t]{0.16\textwidth}
        \centering
        \includegraphics[width=0.5\linewidth]{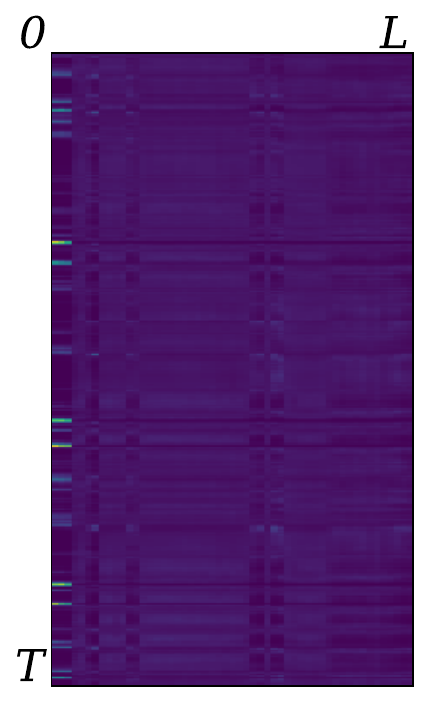}
        \caption{Head 5 of layer 12}
    \end{subfigure}
    \caption{The decoder-only architecture model contains various attention modules with different functionalities that are not always diagonal. We selected a sentence for inference and obtained the attention maps for each head of every layer. In the figure, one head from each decoder layer is randomly selected for illustration.}
    \label{fig:all-attn-maps}
\end{figure*}

With the widespread use of discrete audio tokens \cite{hsu20221hubert,Kumar2023HighFidelityAC,defossez2022high-encodec}, the research paradigm in language models \cite{devlin2018bert,NEURIPS2020_1457c0d6-gpt3} has shown a profound impact on speech modeling and synthesis \cite{lakhotia-etal-2021-generative-glsm,du2022vqtts}. Motivated by recent advancements in auto-regressive (AR) models employing decoder-only architectures for text generation \cite{NEURIPS2020_1457c0d6-gpt3,touvron2023llama}, several studies, such as VALL-E \cite{Wang2023NeuralCL} and BASE TTS \cite{lajszczak2024base} apply similar architectures to TTS task. These studies demonstrate the remarkable capacity of decoder-only architectures in producing natural-sounding speech.
Unfortunately, robustness is a challenging issue for such models, due to the lack of explicit monotonic alignment constraints and the gap between teacher-forced training and AR inference \cite{pmlr-v119-peng20a}. 
To mitigate this, VALL-T \cite{du2024vall} proposes a decoder-only transducer for imposing monotonic alignment constraints. However, this method requires training a new model under a different criterion. Drawing from experience with prior TTS systems, we believe that existing decoder-only TTS models should also implicitly learn some form of alignment via the internal attention black box during training. If we can identify the internal representation of alignment within the model and apply forced alignment constraints during inference, then building a lightweight method to enhance model robustness becomes feasible. A key insight is that in the neural TTS model Tacotron \cite{Wang2017TacotronTE,Shen2017NaturalTS} with an encoder-decoder architecture, researchers discovered diagonal speech-text alignments within its sole attention module's attention map. This suggests that alignments in decoder-only models might also manifest within their attention maps.
However, in decoder-only TTS models that employ massive attentions across various layers and heads \cite{Wang2023NeuralCL,kharitonov2023speartts}, different attention maps are responsible for different functionalities and are not always diagonal \ref{fig:all-attn-maps}. This complicates both awareness and control of the alignment from the attention maps.

\begin{figure*}[t]
    \centering
    \begin{subfigure}[t]{0.24\textwidth}
        \centering
        \includegraphics[width=0.5\linewidth]{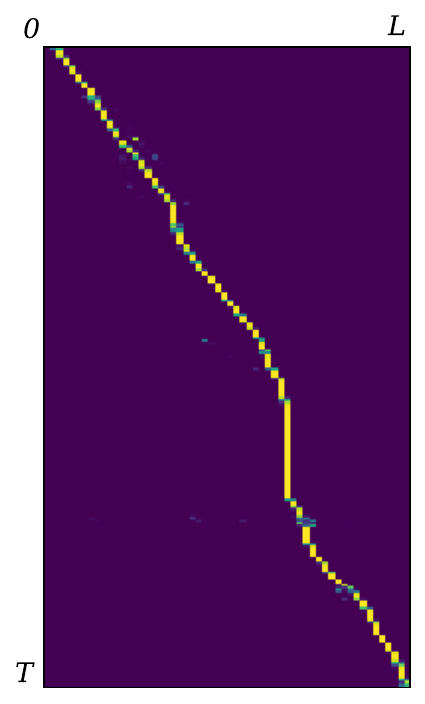}
        \caption{AEAM values for paired data.}
        \label{fig:aeam-value}
    \end{subfigure}
    \hfill
    \begin{subfigure}[t]{0.24\textwidth}
        \centering
        \includegraphics[width=0.5\textwidth]{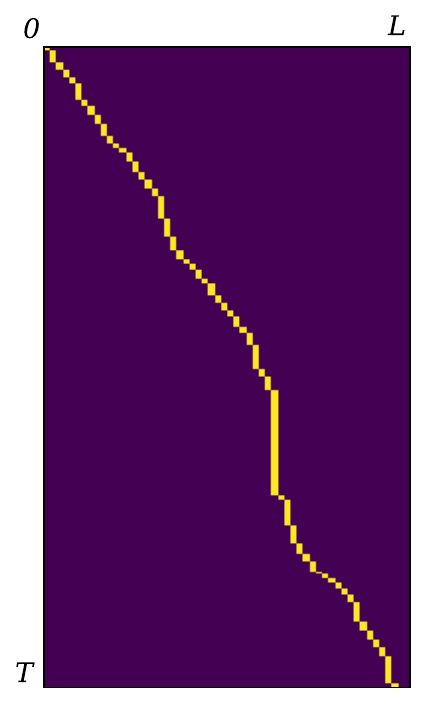}
        \caption{Viterbi-produced alignment.}
        \label{fig:gt-align}
    \end{subfigure}
    % \\
    \begin{subfigure}[t]{0.24\textwidth}
        \centering
        \includegraphics[width=0.5\textwidth]{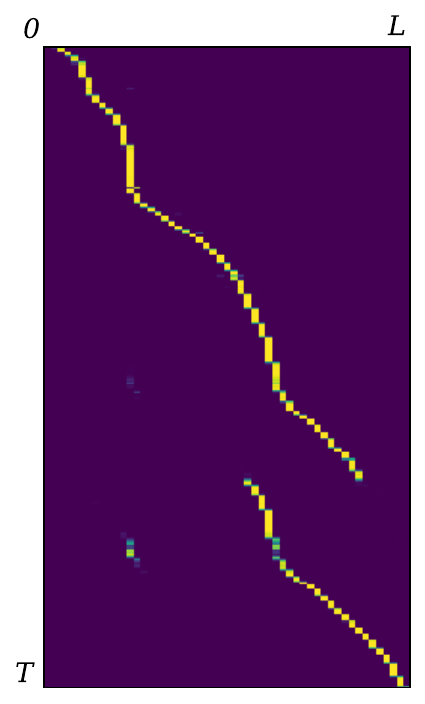}
        \caption{AEAM in a repeat synthesis.}
        \label{fig:word-repeat}
    \end{subfigure}
    \hfill
    \begin{subfigure}[t]{0.24\textwidth}
        \centering
        \includegraphics[width=0.5\textwidth]{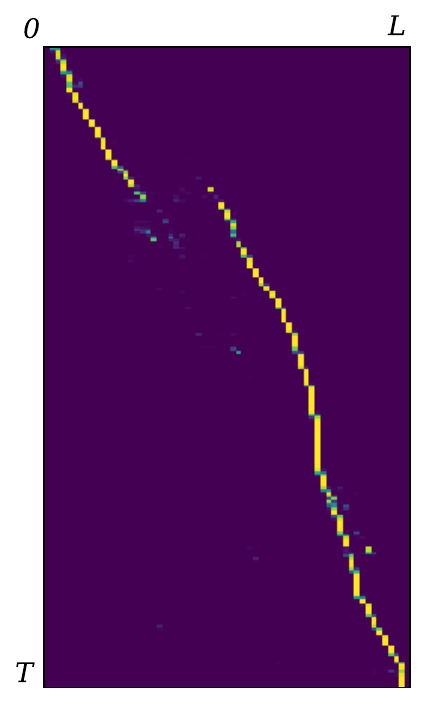}
        \caption{AEAM in a skipping synthesis.}
        \label{fig:word-skip}
    \end{subfigure}
    
    \caption{Illustration of real attention alignments. (a), (c) and (d) are attention heatmaps from the same AEAM of the same model, with different inputs. (a) is for a text-speech pair directly picked from the test set. With the same inputs, (b) shows the reference alignment produced by the Viterbi algorithm. (c) and (d) illustrate the AEAM when some text tokens are repeated or skipped in synthesis.}
    
\end{figure*}

In our initial experiments on the popular decoder-only TTS model VALL-E, we observed from the attention maps that certain attention heads within a specific layer exhibit diagonal patterns, indicating the alignment between speech and text. The attention maps here specifically represent the attention from speech to text, where speech tokens function as queries while text tokens act as keys.
We call the attention maps of those heads Alignment-Emerged Attention Maps (AEAMs, cf. Fig. \ref{fig:aeam-value}). In the circumstances of repeating and skip generation, the alignments on AEAMs also exhibit corresponding patterns (cf. Fig. \ref{fig:word-repeat} and \ref{fig:word-skip}).
Based on this discovery, we introduce a novel inference method named Attention-Constrained Inference (ACI), which facilitates monotonic synthesis without changing the model structure and the training process. It first identifies all AEAMs among the attention maps of all layers and heads using the Attention Sweeping algorithm and then applies constraining masks (CMasks) on AEAMs to guide the monotonic generation process. The main contributions of our work are listed below:

\begin{enumerate}
    \item We discover that among the massive self-attention modules in the whole decoder, some particular heads in a particular early layer are responsible for emerging alignments in their attention maps (i.e., AEAMs). 
    \item We propose a training-free method, Attention-Constrained Inference (ACI), which seamlessly works on top of existing decoder-only TTS models, to detect AEAMs of a model and apply CMasks on AEAMs at the inference stage to guide the alignment to be monotonic and realize robust synthesis.
    \item The word error rate (WER) of synthesized speech is reduced by up to 20.5\% relatively with ACI on VALL-E models of various configurations, while the naturalness and speaker similarity are comparable.
\end{enumerate}

\section{Attention-Constrained Inference}

{In this section, we introduce our Attention-Constrained Inference (ACI) for robust decoder-only TTS, using the Attention Sweeping algorithm to pinpoint AEAMs among all attention maps, and the attention constraining strategies to apply CMasks to AEAMs, helping the model to synthesize monotonically without changing or retraining the model. }

\subsection{Notations}

We denote the discrete tokens of a speech utterance as $\bm{y} = \{y_1, y_2, \dots, y_T\}$ and the transcription text sequence (usually phonemes) as $\bm{x} = \{ x_1, x_2, \dots, x_L\}$. Let the number of layers and heads of VALL-E AR decoder be $N_\mathrm{L}$ and $N_\mathrm{H}$. 
The AR decoder essentially receives a speech sequence and a text sequence to predict the next speech token. 
We further define $\mathcal{E}(\bm{a}, \bm{b}, t) = \frac{1}{t} \sum_{i=1}^{t} (a_i - b_i)^2$, where the numerical sequences $\bm{a}$ and $\bm{b}$ have equal lengths $T$ and $T \ge t$.

\subsection{Attention Sweeping}
We introduce the Attention Sweeping algorithm designed to identify the heads responsible for the emergence of alignments.
This method is initiated by subjecting the AR decoder to conduct a forward pass with several ground-truth text-speech pairs as inputs, aiming to obtain attention maps for every head within all layers. We then analyze these maps to identify AEAMs and record the corresponding heads for subsequent inference. 

An AEAM is expected to manifest the following patterns: 
(1) the attention values of all rows (speech tokens) are concentrated on very few columns (text tokens), and 
(2) as speech tokens are generated, focused text tokens shift towards the sentence end, with the focus positions strongly correlated with the real alignment.
We respectively design the entropy cost and the alignment cost to quantify the extent of conformity an attention map has with the two patterns mentioned above. Note that calculating the alignment cost requires a reference alignment of the input ground-truth utterance, obtainable via a forced alignment algorithm such as Viterbi. However, such reference alignments are solely requisite for the Attention Sweeping stage and are unnecessary for subsequent inference employing attention constraining.
Additionally, before calculating the costs, all attention maps should be normalized such that each row sums up to $1$. The normalized attention map is denoted as $\mathbf{M}^{T\times L}$. 

\textbf{Entropy Cost}.
Since the sum of each row in the normalized matrix \(\mathbf{M}\) equals $1$, we directly use an entropy-like cost to measure the concentration of attention weights in rows of $\mathbf{M}$:
\begin{equation}
    \mathcal{C}_\mathrm{E}(\mathbf{M}) = -\frac{1}{T}\sum_{t=1}^T \sum_{\ell = 1}^L M_{t, \ell} \cdot \log{M_{t, \ell}}.
    \label{eq:entropy-cost}
\end{equation}
Based on the properties of entropy, the more concentrated the attention values are in each row, the lower the entropy cost.

\begin{figure*}[tph]
    \centering
    \begin{subfigure}[t]{0.24\textwidth}
        \centering
        \includegraphics[width=0.79\textwidth]{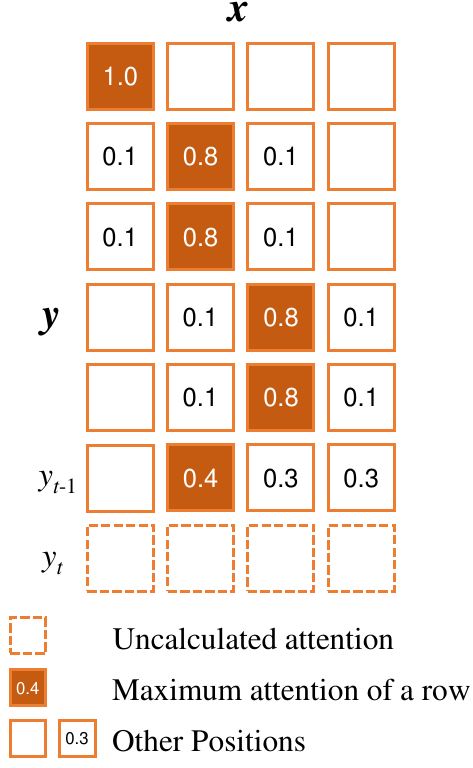}
        \caption{AEAM attention pattern.}
        \label{fig:bot-left-values}
    \end{subfigure}
    \hfill
    \begin{subfigure}[t]{0.24\textwidth}
        \centering
        \includegraphics[width=0.79\textwidth]{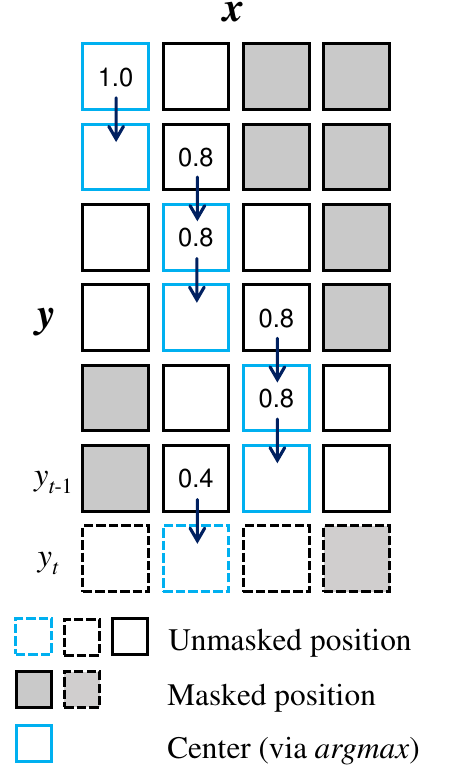}
        \caption{Naive argmax CMask.}
        \label{fig:bot-left-mask-argmax}
    \end{subfigure}
    % \\
    % \vspace{0.5em}
    \begin{subfigure}[t]{0.24\textwidth}
        \centering
        \includegraphics[width=0.79\linewidth]{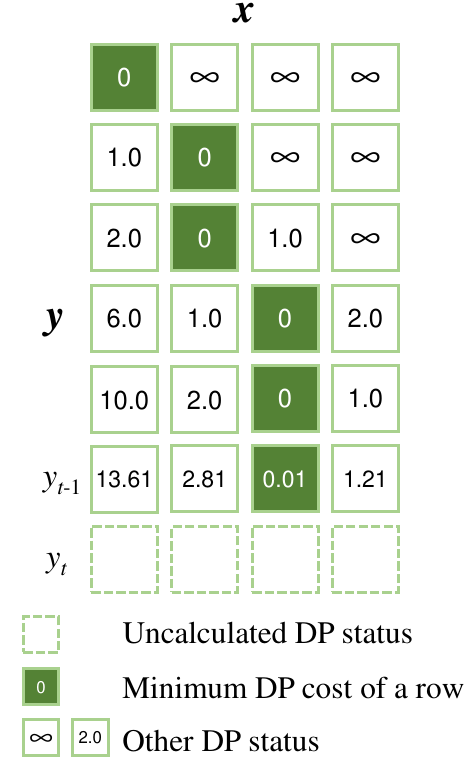}
        \caption{AEAM DP values.}
        \label{fig:bot-left-dp-vales}
    \end{subfigure}
    \hfill
    \begin{subfigure}[t]{0.24\textwidth}
        \centering
        \includegraphics[width=0.79\textwidth]{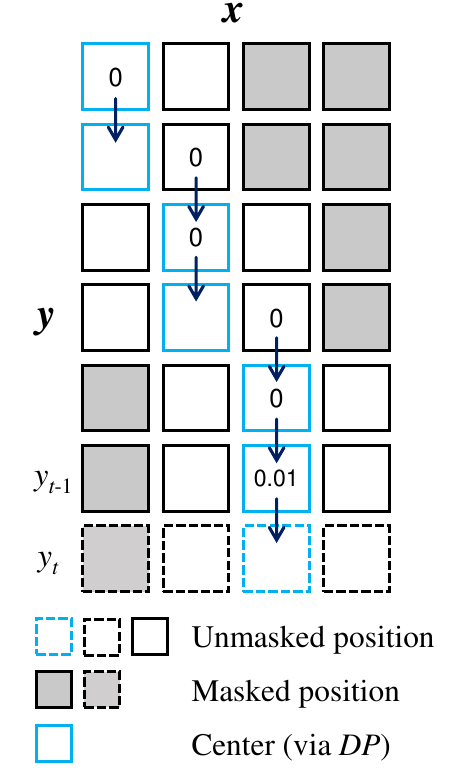}
        \caption{DP CMask.}
        \label{fig:bot-left-mask-dp}
    \end{subfigure}

    \caption{An example of AEAM attention values and DP values, and corresponding CMasks. Here $\bm{x}$ is the input text tokens, and $\bm{y}$ is the speech tokens. (a) presents an example of the attention value pattern of a normalized AEAM when generating $y_{t+1}$, with the maximum value of each row highlighted. The values are normalized so that each row sums up to $1$. The constraining mask (CMask) for the newly generated token $y_t$ is now needed. (b) shows the CMask obtained via the trivial argmax strategy, which is prone to causing instability, such as the attention center being falling back. Based on attention values in (a), subfigure (c) illustrates the DP values $d$ of each state. (d) is the CMask obtained via the DP strategy, which shows stability in maintaining the monotonicity.
}
\end{figure*}

\textbf{Alignment Cost}.  
To further verify whether an attention map conforms to the alignment, we first calculate the average attention position for each row $m_t = \sum_{\ell=1}^{L}{ l \cdot M_{t, \ell} }$. 
Next, we aim to find a sequence $\bm{a}$ that closely matches $\bm{m}$ and satisfies the alignment properties: $\bm{a} = {\arg{\min}}_{\bm{a'}} \mathcal{E}(\bm{m}, \bm{a}', T)$,
where $\bm{a}'$ is a monotonic integer sequence such that $0 \leq a'_{t+1} - a'_t \leq 1$ for $t = 1, 2, \ldots, T-1$, and all its elements fall in range $[1, L]$ (specifically, $a'_1 = 1$ and $a'_T = L$). 
We utilize a dynamic programming (DP) algorithm to determine the sequence $\bm{a}$ that minimizes the cost $\mathcal{E}(\bm{m}, \bm{a}, T)$.
Each potential value for every element of $\bm{a}'$ is treated as a distinct state, so the DP state space contains $T\times L$ states.
We denote $d_{t, \ell}$ as the minimum total cost up to the first $t$ positions (i.e., $\mathcal{E}(\bm{m}, \bm{a}', t)$) when at state $a'_t = \ell$.
When \(t, l \geq 2\), state \(a_t = \ell\) can only be inherited from two prior states: \(a'_{t-1} = \ell\) or \(a'_{t-1} = \ell-1\), so we reach the DP equations (when $t \geq 2$):
\begin{equation}
% \begin{aligned}
d_{t, \ell} = 
\begin{cases} 
d_{t - 1, 1} + (m_t - 1)^2 & \ell = 1, \\
\min \{d_{t - 1, \ell}, d_{t - 1, \ell - 1}\} + (m_t - \ell)^2 &  l \geq 2.
\end{cases}
% \end{aligned}
\label{eq:dp}
\end{equation}
Initial conditions are set at the first row: $d_{1, \ell}$ equals to $(m_1 - 1)^2$ if $\ell = 1$, otherwise $\infty$. Using Eq.~\ref{eq:dp} and calculating \(d_{t, \ell}\) in the order of $t := 1 \to T$, $\ell := 1 \to L$, we can obtain the final minimum total cost \(\min_{\bm{a'}} \mathcal{E}(\bm{m}, \bm{a}', T) = d_{T, L}\), and the state transition sequence for \(d_{T, L}\) is the value assignments for sought sequence \(\bm{a}\). 
Then, we compute the distance between \(\bm{a}\) and the reference alignment \(\bm{b}\), obtained by a forced alignment algorithm like Viterbi. To enhance the tolerance of the detection, we allow a global deviation of a constant \(c\) positions between the two, so the cost becomes \(\min_c \mathcal{E}(\bm{a} + c, \bm{b}, T)\). Now we reach the expression for the alignment cost $\mathcal{C}_{\mathrm{A}}\left(\mathbf{M}\right)$:
\begin{equation}
    \mathcal{C}_{\mathrm{A}}(\mathbf{M}) = \frac{1}{T}\left(\mathcal{E}(\bm{m}, \bm{a}, T) + \min_c \mathcal{E}(\bm{a} + c, \bm{b}, T)\right)
    \label{eq:alignment-cost}
\end{equation}
where $\bm{a}$ is the previously defined monotonic sequence closest to the average attention positions $\bm{m}$. Note that the cost calculation pertains to a single input utterance, but to mitigate instability, we can opt to select multiple utterances and compute the average cost, ensuring a more robust assessment of the correlation of each head to the alignment. Finally, AEAMs can be easily filtered out by setting a threshold $\tau$ on the average of $\mathcal{C}_{\mathrm{E}}$ and $\mathcal{C}_{\mathrm{A}}$: 
\begin{equation}
    \mathcal{C}_{\mathrm{E}}\left(\mathbf{M}\right) + \mathcal{C}_{\mathrm{A}}\left(\mathbf{M}\right) < 2\tau.
    \label{eq:AEAM-thres}
\end{equation}
An attention map is considered an AEAM if and only if Eq.~\ref{eq:AEAM-thres} is satisfied.
In experiments, we found that for models with different configurations and training datasets, after calculating and ranking the entropy cost and alignment cost for all \(N_\mathrm{L} \cdot N_\mathrm{H}\) attention maps, there is always at least one head in a particular early layer (the second or the third layer) whose cost is significantly lower than the others (Section \ref{sec:exp-attn-sweep}). 
An example of heatmap plots for such AEAMs is shown in Fig.~\ref{fig:aeam-value}, clearly illustrating their strong connection with real alignments (Fig.~\ref{fig:gt-align}).

\subsection{Attention Constraining}

VALL-E's AR synthesis often encounters issues with word skips, repeats, or pronunciation errors. We notice that when the text and synthesized speech do not perfectly match, the attention alignment in AEAMs show correspoding patterns (cf. Fig.~\ref{fig:word-repeat} and \ref{fig:word-skip}). 
Assuming that $y_{1}$ to $y_{t}$ have been input or synthesized and that $y_{t+1}$ is to be generated, a possible AEAM value pattern is shown in Fig.~\ref{fig:bot-left-values}. Note that the key in the last row is the newly generated $y_{t}$, for which the alignment position needs to be calculated in this AR iteration. 
However, the AEAM may mistakenly align $y_{t}$ with a phoneme identical to the true one but positioned differently, leading to a non-monotonic attention alignment that misguide the subsequent generation.

\begin{table*}[bt]
    \centering
    \caption{The information on the four VALL-E models used.}
    \centering
    % \scalebox{0.70}{
    \begin{tabular}{c|c c c c c c}
        \hline
        \textbf{Name} & \textbf{\#Params} & $N_\mathrm{L}$ & $N_\mathrm{H}$ & \textbf{Hidden Dim.} & \textbf{Dataset} & \textbf{Epochs} \\ \hline \hline
        {\sc Std}-585h & 367M & 12 & 16 & 1024 & LibriTTS & 40\\
        {\sc Small}-585h & 159M & 9 & 12 & 512 & LibriTTS & 40\\
        {\sc Large}-585h & 1.2B & 18 & 24 & 1536 & LibriTTS & 40\\
        {\sc Std}-5000h & 367M & 12 & 16 & 1024 & LibriLight-6k & 10\\ \hline
    \end{tabular} % }
    \label{tab:model-info}
\end{table*}

To address this, we propose attention constraining strategies, i.e., adding constraining masks (CMasks) on AEAMs to guide the inherent alignment to proceed monotonically. We assign potentially different CMasks to each AEAM based on their values. A straightforward approach is to restrict $y_t$ to only attend to the vicinity of the alignment (i.e., the attention center position) of $y_{t-1}$. The main challenge here is to determine the attention center of $y_{t-1}$ in each AEAM based on the known attention values (i.e., upper $t - 1$ rows of the normalized matrix $\mathbf{M}$). 
A trivial method denoted as the \textit{argmax} strategy (Fig.~\ref{fig:bot-left-mask-argmax}), considers the column with the maximum attention value as the attention center \cite{Tachibana2018efficiently}, but this may lead to instability. As long as the attention values of the preceding token are not concentrated, center fallback is likely to occur. 

Another method denoted as the \textit{DP} strategy (Fig.~\ref{fig:bot-left-mask-dp}), leverages the DP algorithm introduced in Attention Sweeping, selecting the position with the minimum DP cost. Since the global information from the previous $t-1$ rows is taken into account, the \textit{DP} strategy is more stable in maintaining the monotonicity. The two center locating strategies can be summarized as: 
\begin{equation}
\begin{aligned}
    \mathrm{ac}_{t-1} &= {\arg\max}_{\ell} M_{t-1,\ell} \ \ \ \ &\textrm{for } \textit{argmax }\textrm{strategies}, \\
    \mathrm{ac}_{t-1} &= {\arg\min}_{\ell} d_{t-1,\ell} \ \ \ \ &\textrm{for } \textit{DP }\textrm{strategies},
\end{aligned}
\end{equation}
where $\mathrm{ac}_{t}$ is the attention center position for $y_t$. The CMask for $t$-th row is then set as only $[\mathrm{ac}_{t-1} - \rho + 1, \mathrm{ac}_{t-1}+\rho-1]$ unmasked, where $\rho$ is the unmasking radius, set to an integer correlated with the entropy cost of the attention map in experiments. Furthermore, to preserve the original attention values and to simulate monotonic alignments, we respectively design two mask strategies to generate the first new token after the prompt: one that applies the CMask only to the $t$-th row (denoted as the last-row strategy), and another that retains the CMask history for all $t$ rows (denoted as the history-kept strategy). The combination of two center locating strategies and two mask strategies results in four attention constraining strategies, which will be evaluated in the experiments.

\begin{figure*}[tbp]
    \centering
    \begin{subfigure}[t]{0.24\textwidth}
        \centering
        \includegraphics[width=\linewidth]{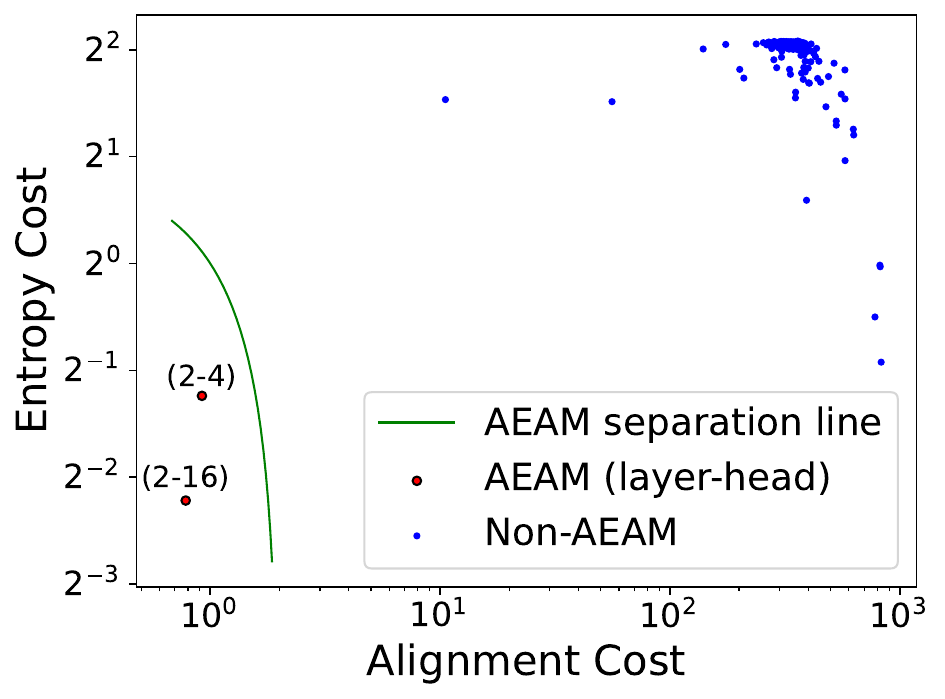}
        
        \caption{{\sc Std}-585h}
        \label{fig:standard-libritts}
    \end{subfigure}
    \hfill
    \begin{subfigure}[t]{0.24\textwidth}
        \centering
        \includegraphics[width=\textwidth]{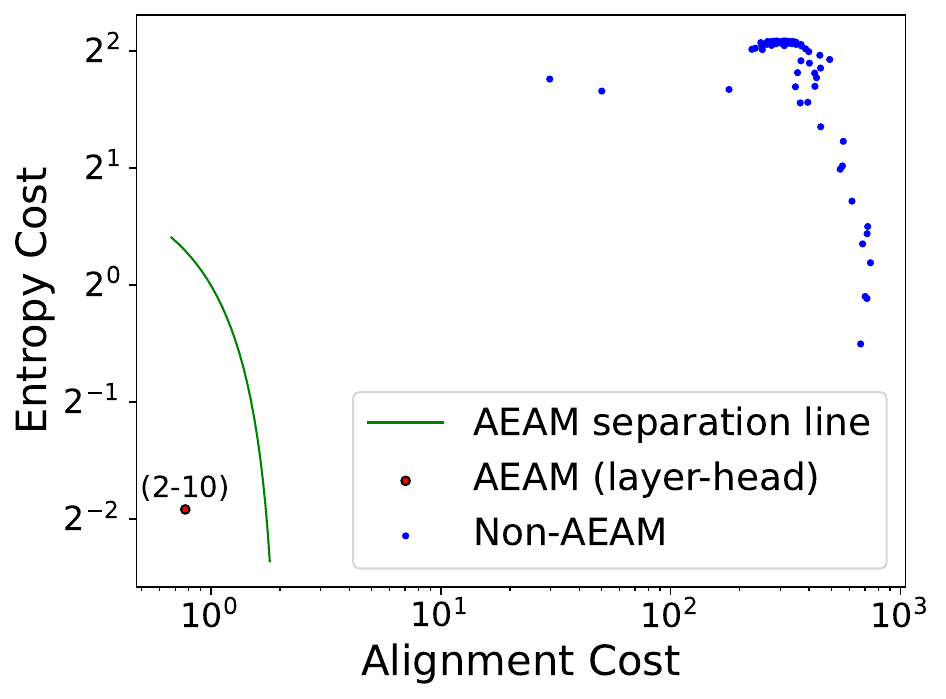}
        
        \caption{{\sc Small}-585h}
        \label{fig:small-libritts}
    \end{subfigure}
    % \\ \vspace{0.5em}
    \begin{subfigure}[t]{0.24\textwidth}
        \centering
        \includegraphics[width=\textwidth]{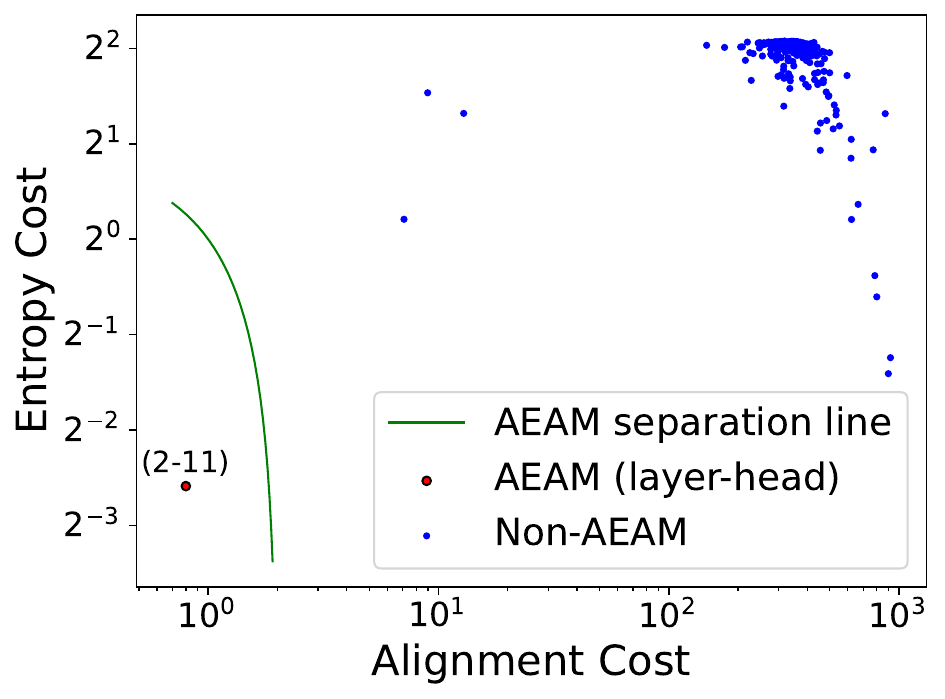}
        
        \caption{{\sc Large}-585h}
        \label{fig:large-libritts}
    \end{subfigure}
    \hfill
    \begin{subfigure}[t]{0.24\textwidth}
        \centering
        \includegraphics[width=\textwidth]{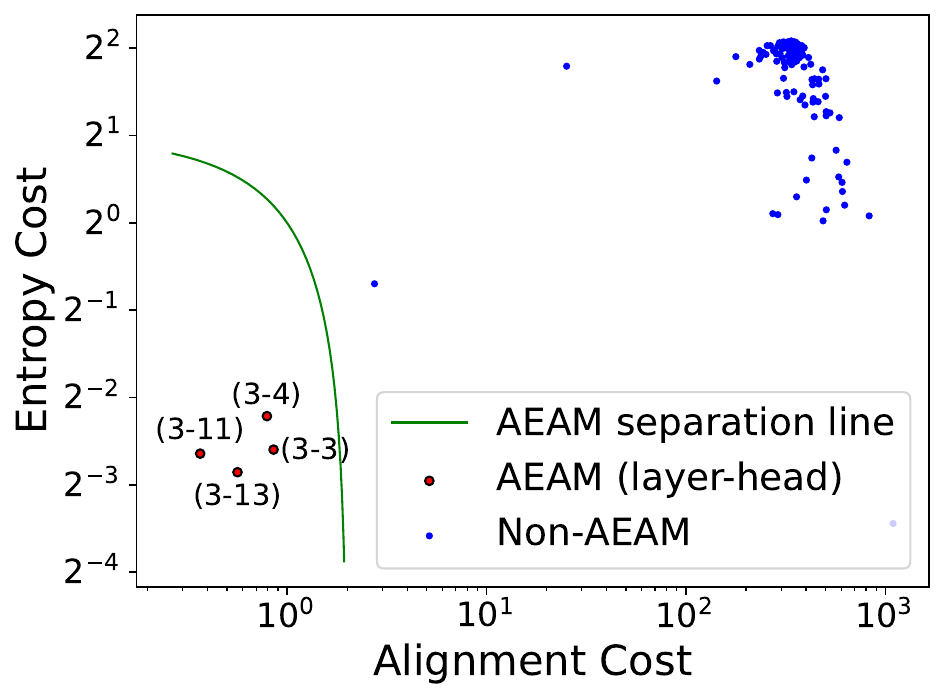}
        
        \caption{{\sc Std}-5000h}
        \label{fig:standard-librilight}
    \end{subfigure}
    
    \caption{Calculate $\mathcal{C}_\mathrm{E}$ and $\mathcal{C}_\mathrm{A}$ for each attention map of 4 models, and plot them as 2-D points. The separation line is set at \(\tau=1\).}
    
    \label{fig:attn-sweep-res}
\end{figure*}

\begin{table*}[tpb]
\centering
\begin{minipage}{1.0\textwidth}
    \centering
    % \begin{minipage}{0.65\textwidth}
    \caption{WER Results for all four models and four attention constraining strategies.}
    \label{tab:wer-results}
    % \scalebox{0.72}{
    \begin{tabular}{c|c>{\centering\arraybackslash}m{21pt}>{\centering\arraybackslash}m{20pt}>{\centering\arraybackslash}m{25pt}|c|c|c}
    \hline
    \multirow{2}{*}{\textbf{Strategy}} 
    & \multicolumn{4}{c|}{{\sc Std}-585h} 
    & \multicolumn{1}{c|}{{\sc Small}-585h} 
    & \multicolumn{1}{c|}{{\sc Large}-585h} 
    & \multicolumn{1}{c}{{\sc Std}-5000h} \\ \cline{2-8}
    & \textbf{WER}(\textbf{\%}) $\downarrow$   & Del(\%) 
    & Ins(\%)   & Sub(\%) 
    & \textbf{WER}(\textbf{\%}) $\downarrow$   & \textbf{WER}(\textbf{\%}) $\downarrow$   
    & \textbf{WER}(\textbf{\%}) $\downarrow$       \\ \hline \hline
    None & 4.78 & 0.74 & 0.84 & 3.20 & 6.65 & 4.60 & 4.31  \\ \hline
    \textit{argmax}/last-row & 4.46 & 0.65  & 0.67 & 3.13 & 5.63 & 4.13 & 3.62        \\
    \textit{argmax}/history-kept & 6.68 & 0.82 & 1.73 & 4.13 & 6.25 & 4.99 & 3.88     \\
    \textit{DP}/last-row & 4.26 & \textbf{0.64} & 0.79& \textbf{2.82} & 5.56 & 4.12 & \textbf{3.55}  \\
    \textit{DP}/history-kept & \textbf{4.14}& 0.65  & \textbf{0.65}& 2.84  & \textbf{5.29}& \textbf{3.95}  & 3.57    \\ \hline
    \end{tabular}%}
\end{minipage}
\\
\begin{minipage}{1.0\textwidth}
    \centering
    \caption{MCD, Predicted-MOS (P-MOS), and SECS for {\sc Std}-585h model and DP strategies.}
    \label{tab:mcd-mos-secs}
    \centering
    % \scalebox{0.72}{
    \begin{tabular}{c|ccc}
    \hline
    {\textbf{Strategy} (DP)} 
    & \textbf{MCD} $\downarrow$   & \textbf{P-MOS} $\uparrow$ & \textbf{SECS} $\uparrow$  \\ \hline \hline
    None         & 3.94          & 4.40 & 0.856\\ \hline
    last-row     & 3.93          & 4.40 & 0.859\\ 
    history-kept & \textbf{3.92} & \textbf{4.41} &\textbf{0.860} \\ \hline
    \end{tabular}%}
\end{minipage}
\\
\begin{minipage}{1.0\textwidth}
    \centering
    \caption{{\sc Std}-585h model and \textit{DP}/history-kept strategy: impact of applying constraining masks on less or more AEAMs}
    \centering
    % \scalebox{0.77}{
    \begin{tabular}{c|cccc}
    \hline
    \textbf{Attn. Constrained AEAM(s)} 
    & \textbf{WER}(\textbf{\%}) $\downarrow$   & Del(\%) & Ins(\%) & Sub(\%)  \\ \hline \hline
    (2-16) and (2-4)     & \textbf{4.14} & \textbf{0.65} & \textbf{0.65} & \textbf{2.84}\\ 
    (2-16)               & 4.28 & 0.68 & 0.71 & 2.89 \\ 
    all 2-nd layer attention maps   & 4.76 & 0.89 & 0.80 & 3.06 \\ \hline
    \end{tabular}%}
    \label{tab:ablation}
\end{minipage}
\end{table*}

\section{Experiment}

\subsection{Setup}

We use Encodec \cite{defossez2022high-encodec} tokenizer with a \SI{16000}{\hertz} sampling rate and a  \SI{20}{\milli\second} frameshift to quantize each speech frame into 8 RVQ indices. We use a widely recognized implementation of VALL-E \footnote{https://github.com/lifeiteng/vall-e} as the base code. We then train four VALL-E models with three different sizes and on two different multi-speaker datasets (LibriTTS \cite{zen2019libritts} with a total of 585 hours and 2306 speakers, and LibriLight-6k \cite{kahn2020libri} with transcribed text in LibriHeavy \cite{kang2024libriheavy} (with a total of 5000 hours and 1531 speakers), whose detailed information is shown in Table~\ref{tab:model-info}.  With each model configuration, both AR and on-AR stages are trained. We use the ScaledAdam \cite{yao2024zipformer} optimizer and the Eden \cite{yao2024zipformer} scheduler with a 0.05 base learning rate to train all models. 

\subsection{Attention Sweeping}
\label{sec:exp-attn-sweep}

We first use the Attention Sweeping algorithm to filter out the potential AEAMs. For the four models in Table~\ref{tab:model-info}, we respectively conduct the algorithm to sweep all $N_\mathrm{L} \cdot N_\mathrm{H}$ attention maps. We randomly select 5 text-speech pairs from the LibriTTS validation set, feeding them into the models to obtain the attention map values. For each attention map, we follow Eq.~\ref{eq:entropy-cost} and Eq.~\ref{eq:alignment-cost} to calculate its entropy cost $\mathcal{C}_\mathrm{E}$ and alignment cost $\mathcal{C}_\mathrm{A}$. The reference alignments are obtained via conducting the forced alignment algorithm in Kaldi \cite{Povey2011TheKS}, also employing a frameshift of \SI{20}{\milli\second}.
Then, we average the cost of 5 utterances, so that each attention map can be regarded as a point on a 2-D plot whose y-axis is for $\mathcal{C}_\mathrm{E}$ and x-axis is for $\mathcal{C}_\mathrm{A}$. Simply setting the AEAM threshold $\tau$ (i.e., the average of $\mathcal{C}_\mathrm{E}$ and $\mathcal{C}_\mathrm{A}$, see Eq.~\ref{eq:AEAM-thres}) as 1, we obtain the results of Attention Sweeping, shown in Fig.~\ref{fig:attn-sweep-res}. 
The y-axis and x-axis of the plot are log-scaled with bases 2 and 10, respectively. Points located below the AEAM separation line $\tau = 1$ (green line) represent attention maps considered as AEAMs. The four models each have between one to four AEAMs, with their attention value distributions being concentrated and strongly correlated with the true alignment (see Fig.~\ref{fig:aeam-value} and \ref{fig:gt-align}). 
Additionally, note that each model's AEAMs are located within the same Transformer layer, specifically in the relatively early second or third layer. This indicates that VALL-E treats the text-speech alignment as a shallow-level feature of the text-speech joint distribution, which implies that by constraining these shallow-layer attentions, we can influence the inference of deeper layers to improve the robustness.

\subsection{Robust Generation}

\label{sec:exp-robust-gen}

Following the identification of AEAMs and the corresponding layers and heads for each model, we can enhance the robustness of synthesis using attention constraining strategies. Our test set, mirroring the one employed in UniCATS \cite{Du2023UniCATSAU}, includes 500 utterances from 37 speakers in the LibriTTS test set, with each speaker assigned a distinct speech prompt. We use the word error rate (WER) as our main metric to measure the robustness of zero-shot TTS synthesis. 
Our evaluation process for each VALL-E model and attention constraining strategy involves performing inference on the test set (applying the strategy during the AR stage and leaving the non-AR stage untouched), then reconstructing the RVQ indices into speech, and finally transcribing the speech into text using the ASR model, Whisper\footnote{https://huggingface.co/openai/whisper-medium} \cite{radford2023robust}. The WER is then calculated by comparing the ground-truth text and the transcribing text. 
The CMask radius $\rho$ for attention map $\mathbf{M}$ is set to $\mathrm{round}\left(8\mathcal{C}_\mathrm{E}\left(\mathbf{M}\right)\right) + 1$, and the baseline is not to employ any attention constraining strategies, i.e., the original VALL-E inference. The results, listed in Table~\ref{tab:wer-results}, show that while the \textit{argmax} strategies sometimes perform unstably, the \textit{DP} strategies obtain a WER reduction of 13.4\% to 20.5\% in each model, and the counts of deletions, insertions, and substitutions are all lower than the baseline. 

Additionally, we evaluated the mel-cepstral distortion (MCD) generated under the DP strategies to measure the distance between generated speech and ground truth. Pre-trained NISQA\footnote{https://github.com/gabrielmittag/NISQA} \cite{mittag2021nisqa} and Resemblyzer\footnote{https://github.com/resemble-ai/Resemblyzer} models are also utilized to assess Predicted Mean Opinion Score (P-MOS) and Speaker Embedding Cosine Similarity (SECS), respectively, evaluating the synthesis naturalness (automatically predicted by the NISQA model) and the speaker similarity compared to the prompt. Results in Table~\ref{tab:mcd-mos-secs} indicate that the speech synthesized using ACI shows no degradation in the aforementioned metrics. Subjective auditory perception also indicates that the naturalness under DP strategies is slightly higher or on par with the baseline.

\subsection{Ablation Study}
The strategy employed in Section~\ref{sec:exp-robust-gen} involved applying CMasks across all AEAMs. This section conducts an ablation study where, during the inference of the {\sc Std}-585h model using \textit{DP} with history kept, only the head with the lowest average cost, specifically the 16-th head of the 2-nd layer (denoted as (2-16)), is subjected to a CMask. We also tested the results of applying CMasks to all 2-nd layer attention maps. The results, as shown in Table~\ref{tab:ablation}, indicate that only applying CMasks to all AEAMs below the $\tau=1$ line is the most robust choice.

\section{Conclusions}

Our paper introduced a training-free method, ACI, to improve the decoder-only TTS model VALL-E's synthesis robustness. Utilizing Attention Sweeping, we identify and filter AEAMs, strongly related to the alignment, then employ attention constraining to mitigate synthesis errors. Experimental results demonstrate noticeable improvements in WER without degrading naturalness. 
Our work demonstrates that decoder-only TTS models can unsupervisedly learn alignment-like structures and shows a way to locate and exploit them, laying the foundation for various potential works, such as streaming generation or integrate additional signals into attention maps during training.

\newpage
\bibliographystyle{IEEEbib}
\bibliography{mybib}

\end{document}